# A Brief Study of Privacy-Preserving Practices (PPP) in Data Mining

Dhinakaran D[1], Joe Prathap P.M[2]

*Assistant Professor, Department of Information Technology, Velammal Institute of Technology, Panchetti, Tiruvallur, India[1]*

*Associate Professor, Department of Information Technology, RMD Engineering College, Kavaraipettai, Tiruvallur, India[2]*

Abstract:
Data mining is the way toward mining fascinating patterns or information from an enormous level of the database. Data mining additionally opens another risk to privacy and data security. One of the maximum significant themes in the research fieldis privacy-preserving DM (PPDM). Along these lines, the investigation of ensuring delicate information and securing sensitive mined snippets of data without yielding the utility of the information in a dispersed domain. Extracted information from the analysis can be rules, clusters, meaningful patterns, trends or classification models. Privacy breach occur at some stage in the communication of data and aggregation of data. So far, many effective methods and techniques have been developed for privacy-preserving data mining, but yields into information loss and side effects on data utility and data mining effectiveness downgraded. In the focal point of consideration on the viability of Data Mining, Privacy and rightness should be improved and to lessen the expense.

## I. INTRODUCTION

Data miningis one of the quickly expanding fields in the PC business that manages to find valuable and fascinating examples covered up in tremendous measures of information put away in various data sources [1]. Data mining is aversatilefield uniting strategy from Database innovation, Statistics, Information recovery (IR), AI, and Machine learning, Pattern acknowledgment, Neural systems, Knowledge-based frameworks, High-execution computing, and Data perception to report the issue of data. It is used to extract valuable information for future predicting and improvement [2].

Data mining plays an essential role in different business associations, monetary, instructive and wellbeing associations and uncovering sensitive information from there data, which might be large harm if the information known to the outsider.

From the perspective of the association, mining is useful for their future determining and upgrade. In this way, there is a need to avoid the revelation of private data and realize which is viewed as touchy in some random setting. Because of this, numerous endeavors have been devoted to attention to the issue of privacy preserving in data mining [3]. Thus, a few privacy-preserving techniques join with protection safeguarding systems has been created.

Concerns of Data Mining

A. Privacy concern

Privacy-focused on confidentiality, where information about the individuals are not publicized to others [4]. hence providing less information regarding individual users while learning more quantity of information regarding statistical information. where information proprietors like clients, representatives, internet-based life clients are frightened to give their data because of an unapproved individual mayget to their sensitive information and utilize that information in a deceptive manner which may make hurt them.Privacy realized primarily by Cryptography, Anonymization, obfuscation and differential privacy.

1. Cryptography

By using cryptographic algorithms, the data are transformed from one form to another, thus transformed data will be exact and protected. Cryptography has a limitation, where it fails when multiple parties are involved.

2. Anonymization

Anonymizationinvolves either encryption or removal of an individual's identifiable information. limitations of Anonymization are heavy loss of pieces of information and the heavy possibility of linking attacks [5].

3. Obfuscation

Obfuscation is a procedure that modifies the information so as to shroud data. It is a type of information refinement which renovate individual's information into imprecise information making it intricate to recognize. limitations of obfuscation are, loss of individual's information and reverse-engineering a program will be intricate and protracted, it will not essentially make it impossible.

4. Differential Privacy

It includes that calculations be inhumane toward changes in a specific person's record, in this way limiting information spills through the outcomes. The protection saving interface guarantees unequivocally safe access to the information and doesn't require from the information excavator any skill insecurity. utilizing differential protection procedures, we can augment the precision of questions from measurable data and decrease the likelihood of distinguishing a person's data.

B. Security Concern

Security is to keep unapproved clients from picking up anything about the original data.Organizations have lots of personal information's about employees, customers, patients, etc. They don't have adequate security frameworks set up to ensure this data. In heaps of situations where hackers access and took individual information of clients. Security is guaranteed principally through access control components, For example, confirmation and approval [6]. Security can't put a stop to protection divulgence.For example, if one of the users decrypts the datasets, hence entire datasets will be visible which impacts to loss of privacy.

## II. PRIVACY PRESERVING DATA MINING

The study of accomplishing data mining objectives starved oftrailing the privacy of the data owners. If there is a huge amount of data means that it is possible to study a bunch of information about individuals from a group of people's data. Privacy-Preserving DM manages ensuring the protection of individual information/delicate information without relinquishing the utility of the original data [3].

For example, suppose different sets of Hospitals wish to identify valuable summative information or knowledge about the specific disease from their patient's report, while all hospitals will be not set to exposepatient's data in light of security acts.Hence, they have to depend on the privacy mechanism on their distributed database to achieve the needed information.PPDM started with the exertion of Agrawal and Srikant, which accomplished prevalence in the information mining research network

A. Aim of PPDM algorithm

Peoples are well conscious of privacy invasions on their sensitive data and they are very disinclined to share sensitive pieces of information with others. Hence theaim of PPDM is to,

- Preserve privacy of the party's sensitive data. while they achieve useful information from a complete dataset.
- To determine valuable information from available Sanitized data.
- Must be imperviousto the several DM techniques.
- Should not conciliation the access and the utilize of non-sensitive data.
- To allow the data excavator to get perfect mining outcomes by not provided with original data.

B. Common Approaches to protect privacy

- Restrict entree to the records, such as adding up authentication like certification to data entries.
- Replace individuality with pseudonyms or special cryptogram, hence sensitive information can't be located to an individual record.
- Lindell and Pinkas (2009) introduced a concept where data is scattered among several locations and these locations cooperate to study the universal data mining outcomes without disclosing the data at their distinct locations [7].
- By simply applying data mining techniques over multiple sources independently at each place that will never share their information and finally combining the mined results. This concept yields better results locally but fails to give an accurate result globally.

C. Privacy Preserving structure:

Data collected from each individual user (database / data marts) are aggregated and placed in a common database. Then data transformation / sanitization process like blocking, suppression, perturbation, modification, generalization, sampling is done, where data are transformed to the format suits for analysis purpose. Thus, sensible data's will be not exposed even to dishonest data miners.At last data mining algorithms are applied on the transformed database to generate the valuable information.

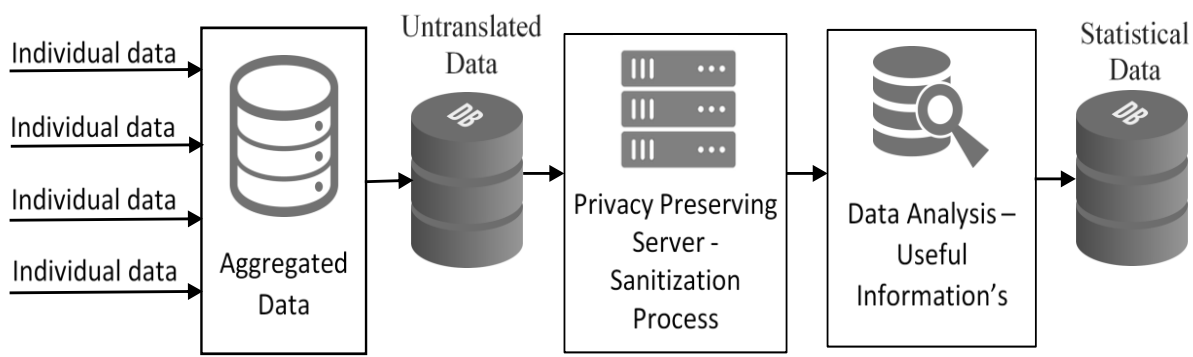

Fig. 1. Privacy Preserving structure – Trusted Process

Trusted: Privacy Preserving process espouse the principles of broadcast, purification, examination and deploying the rules, patterns or end results.

Fig. 1. Shows the trusted process of secrecy preserving.

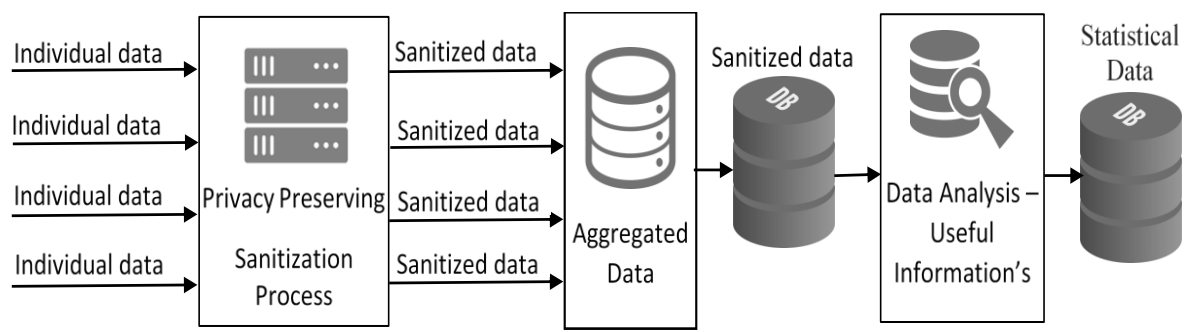

Fig. 2. Privacy Preserving structure –Un-trusted Process

Un-trusted: Privacy Preserving process espouse the principles of purification, broadcast, examination and deploying the rules, patterns or end results. Fig. 2. Shows the un-trusted process of secrecy preserving.

D. Privacy Attacks

1. Linking attack

Realizing of a person through a range of fields or characteristics, for example, the mix of postal district, age, sex, or a few unimportant data. An intruder can coordinate anonymized information with non-anonymized information in an alternate dataset, prompting security break.

2. Homogeneity Attack

This assault uses the situation where everyone of the qualities for a touchy incentive inside a lot of k records is indistinguishable. In such property, although the information has been k-anonymized, the delicate incentive for the arrangement of k records might be anticipated.

3. Background Knowledge Attack

This assault uses a relationship between at least one semi identifier properties with the touchy ascribe to diminish the arrangement of potential qualities for the delicate characteristic.

4. Deduction Attack

The assault can utilize a few information mining procedures, for example, association rules and Bayesian reasoning, to dispatch an assault to misguidedly pick up private information about people by getting to some open data or the yield of some calculation depending on the individual data of people.

5. Correlation Attack

An assault with Background Knowledge or assistant data on co relational data has a high possibility of acquiring privacy data, consequently disregarding privacy.

## III. CLASSIFICATION OF PPDM TECHNIQUES

A. Data Distribution

Data Distribution is a process where data are detached as horizontally or vertically.

1. Horizontally partitioned data

Involves placing different rows/tuples of the database into diverse tables. Perhaps people with age fewer than 18 are stored in Children's, while people with age superior than or identical to 18 are stored in Adults. The two partition tables are then Children's and Adults, while a view with a union might be created over mutually to provide a complete view of all peoples.

2. Vertically partitioned data: involves the creation of tables with aentity of selected attributes/columns from the unique table and using additional tables to store the remaining attributes/columns of the unique table [8]. Normalization likewise includes this parting of sections crosswise over tables, yet vertical apportioning goes more distant than that and parcels segments in any event, when recently standardized.

B. Data Modification

Altering the data commenced in the database using various methods.

1. Perturbation – substituting the columndatacommenced in the table by a differentdata. For example, substituting zero with one and one with zero, this process is said to be adding noise.
2. Blocking - substituting the column data from the table by a special symbol "?".
3. Swapping – Altering the column data from the table.
4. Sampling – Selecting only the sample of data from the data pool.
5. Encryption – Converting plain data into Cipher data using various Cryptographic techniques.

C. Data mining algorithm

The Third aspect of Classification of PPDM Tech is data mining algorithms that are applied to alter data to change information to get the helpful chunk of data that was enclosed up ahead of time. The mining calculations incorporate classification mining, association rule mining, clustering, and Bayesian networks and so on.

D. Data or Rule Hiding

Data - Protecting sensitive data values by concealing the data commenced in the data pool. Delicate data like name, personality, and address from the first dataset that can be connected, straightforwardly or in a roundabout way, to a distinct individual are covered up.

Rule – Protecting Confidential patterns or information derived from data analysis by hiding its rule.

E. Privacy Preservation

Itadverts to the procedures that are used to protect privacy. Furthermore, to save security, information change ought to be done cautiously to accomplish high information utility.Privacy Preservationclassified in to five categories.

- Anonymization
- Perturbation
- Cryptographic based PPDM
- Randomized response based PPDM
- Condensation Based

**IV. PPDM TECHNIQUES**

A. Anonymization

Anonymizationrefers to classify the sensitive information of the data-owner and hiding it. Privacy is questionable if quasi-identifiers are associated to widely existing data, such attacks are known as linking attacks. Linking attacks can direct to predicting with a higher probability of an individual's data [9].Common anonymization approaches are: Generalization, Suppression, swapping, Bucketization and Randomization. Fig.3. provides a clear examination ofanonymization technique, where attribute name detached ahead.

1. Generalization

Substitute value by a lesser amount of explicit yet semantically unswerving value [9], [10]. For example, the age of cashier is 21 from the displayed table, as an outcome of generalization age will be generalized into 15 – 25, which is semantically reliable. Various forms of generalization are Full-domain generalization, Sibling generalization,cell generalization, Sub-tree generalization, and multidimensional generalization.

| Gender | Age | Department | Designation | Salary |
|---|---|---|---|---|
| Male | 21 | Sales | Cashier | 23,430 |
| Female | 24 | Merchandise | Merchandiser | 32,056 |
| Male | 43 | Sales | Sales | 42,72 |

| Gender | Age | Department | Designation | Salary |
|---|---|---|---|---|
| At all | 15-25 | Sales | Cashier | XX,430 |
| At all | 15-25 | Merchandise | Merchandiser | XX,056 |
| At all | 35- | Sales | Sales | XX,72 |

| Gender | Age | Department | Designation | Salary |
|---|---|---|---|---|
|  |  |  | Associate | 9 |
| Female | 31 | Logistics | Stocking Associate | 37,482 |
| Male | 27 | Merchandise | Merchandiser | 38,800 |
| Male | 54 | Marketing | Manager | 45,030 |
| Female | 37 | Logistics | Stocking Associate | 44,270 |

Original Data

| Gender | Age | Department | Designation | Salary |
|---|---|---|---|---|
|  | 45 |  | Associate | 9 |
| At all | 25-35 | Logistics | Stocking Associate | XX,482 |
| At all | 25-35 | Merchandise | Merchandiser | XX,800 |
| At all | 45-55 | Marketing | Manager | XX,030 |
| At all | 33-45 | Logistics | Stocking Associate | XX,270 |

Generalized Data

| Gender | Age | Department | Designation | Salary |
|---|---|---|---|---|
| Male | ## | Sales | Cashier | ##### |
| Female | ## | Merchandise | Merchandiser | ##### |
| Male | ## | Sales | Sales Associate | ##### |
| Female | ## | Logistics | Stocking Associate | ##### |
| Male | ## | Merchandise | Merchandiser | ##### |
| Male | ## | Marketing | Manager | ##### |
| Female | ## | Logistics | Stocking Associate | ##### |

Suppressed Data List

| Gender | Age | Department | Designation | Salary |
|---|---|---|---|---|
| Male | 21 | Sales | Cashier | 23,430 |
| Female | 37 | Logistics | Stocking Associate | 44,270 |
| Male | 27 | Merchandise | Merchandiser | 38,800 |
| Female | 31 | Logistics | Stocking Associate | 37,482 |
| Male | 54 | Marketing | Manager | 45,030 |
| Male | 43 | Sales | Sales Associate | 42,729 |
| Female | 24 | Merchandise | Merchandiser | 32,056 |

Swapped Data List

| Gender | Age | Department | Designation | Bucket ID |
|---|---|---|---|---|
| Male | 21 | Sales | Cashier | B1 |
| Female | 24 | Merchandise | Merchandiser | B1 |
| Male | 43 | Sales | Sales Associate | B1 |
| Female | 31 | Logistics | Stocking Associate | B1 |
| Male | 27 | Merchandise | Merchandiser | B2 |
| Male | 54 | Marketing | Manager | B2 |
| Female | 37 | Logistics | Stocking Associate | B2 |

| Bucket ID | Salary |
|---|---|
| B1 | 23,430 |
| B1 | 32,056 |
| B1 | 42,729 |
| B1 | 37,482 |
| B2 | 38,800 |
| B2 | 45,030 |
| B2 | 44,270 |

Bucketized List of Data

Fig. 3. Various Anonymization Approaches

2. Suppression

Engage with jamming the values with special symbols, representing that the value has been stifled [11]. For example, the age of cashier is 21 from the displayed table, as aresult of the Suppression technique, age will be Suppressed to ##, which the value is jammed with distinctsigns ##. Various forms of Suppression techniques are rounding, generalization, using intervals.

3. Swapping

Trade of information which might have sensitive attributes among the two posse [12]. Example, Exchanging the X tuple sensitive attribute – Salary by Y tuple. This technique gives better outcomes, meaningful information and very effective for huge datasets.

4. Bucketization

Bucketization is the method of dividing the primary data-table into various buckets, one of thebucketscomprisejust sensitive attributes and some other bucket contains the rest of the attributes of the original data table. Example, in the, showed table, salary is assumed to be a sensitive attribute, consequently partitioning that attribute and putting in a separate bucket and rest of the attributes correspondingto age, gender, department, and designation placed in the independent table

5. Randomization

Randomization process is a technique in the direction of irritating the information to the distributed data mining technique consequently that the information approximations of perceptive components are protected from the recognition [13] – [15]. Sufficient noise is appended in original records to protect from the recovery. The altered data is indistinguishable with the primary data. Various forms of randomization are Including Random numbers, Creating random vectors, Random change of a sequence.

Sweeney (2002)proposed K-anonymity using Suppression and Generalization to attain K-anonymity. where each individual data is distinct from others as a minimum k-1. Emancipating such data for analysis will decrease the take a chance of identification when collective with visibly available data.Many innovative approaches have been proposed, such as p-sensitive k-anonymity,t-closeness, (a, k)-anonymity,M-invariance,I-diversity and personalized anonymity [26].

Limitations

Accurateness of the data-analysis on altered data is reduced. Hence this methodgrieves from heavy information loss. Homogeneousness and contextual knowledge attack also give way to the expose of individual's data. Additional two major downsides are, First It could be extremely intense for the proprietor of the database to figure out which of the properties are reachable and which are absent in outside tables. Second, k-obscurity model thinks about a specific approach of assault, while in genuine circumstances there is not any legitimization that why the aggressor ought not attempt different approaches.

B. Perturbation

The actual record values are supplanted with few roughly artificial data values thus, statistical information evaluated from the annoyed factsdoesn't contrast from the statistical information figured from the actual data.Intruder can't accomplish linking attack or convalesce sensitive facts from the obtained data.The annoyed data records don't correspond to true record proprietors; in this way, an aggressor can't get back sensitive data from the distributed information or play out the delicate linkages. Since just factual properties of the records are held, in this way, bothered information which contains the individual records are useless to the beneficiary.Perturbation can be achieved by adding noise, data swapping, artificial data generation.Perturbation technique treats dissimilar attributes autonomously.Since in perturbation system, as conflicting to remaking the first qualities just the disseminations are remade. Thus, new calculations ought to be created which use these remade dispersions to execute mining on the information. In this way, along these lines for every individual information issue, for

example,association rule mining, clustering, or classification,another appropriation-based information mining algorithm ought to be created.

Limitations

- Loss of hidden statistics existing in multidimensional data in distributed database system.
- Original data values can't be reconstructed.

C. Cryptographic based PPDM

When several persons work organized to calculate common results and thus keep away from expose of sensitive information. The gatherings associated with such kind of assignment might be contenders or not confided in parties, so protection turns into the primary concern. Cryptographic systems are ideally inferred in such circumstances where various parties work mutually to survey results or offer non-sensitive mining results and in this way abstaining from uncovering of touchy data [17]. Cryptographic procedures discover its handiness in such circumstances as a result of two reasons: First, it gives a well-characterized model to security that includes techniques for computing and demonstrating it. Second, an expansive transcription of cryptographic computationis accessible to portray security saving information mining computation in this area.

The data might be dispersed either horizontally or vertically. Vaidya and Clifton built up a Naive Bayes classifier for safeguarding protection on vertically divided information. Vaidya and Clifton in proposed a technique for grouping over vertically parceled information [18]. Every one of these techniques depend on an encryption convention is known as Secure Multiparty Computation (SMC) innovation. SMC characterizes two essential antagonistic models in particular (I) Semi-Honest model and (ii) Malicious model. Semi-honest model pursues conventions sincerely however can attempt to reason the mystery data of different gatherings. In the Malicious model, pernicious foes can successfully construe mystery data. There exist many numbers of solutions in case of Semi-Honest model, but in case of Malicious model there are extremely fewer studies have been made.Without a doubt, this methodology guarantees that changed information is precise and secure yet this methodology bombs when multiple gatherings are included. Likewise, the last mining outcomes may lead to protection loss of individual records. One of the applications of cryptography is, highly secure online auctions management system[6].

Merits

- Can manage disadvantages of perturbation technique mentioned above.
- Altered data are accurate and protected.

Limitation

- Where it fails when multiple parties are involved.
- It is much low efficient than other techniques.

D. Randomized response based PPDM

This statistical technique has been established by Warner, which typically utilized in the perspective of altering data by probability distribution for methods such as a survey [13] – [15].Two models: Related-Question Model and Unrelated-Question Model have been createdasthe solution of survey problem. In the Related-Question Model, instead ofquerying every individual whether they need to property A the questioner asks every individual two related inquiries, the responses to which are restricting one another. The procedure of assortment of information in Randomized reaction method is completed in two stages

- Information suppliers randomize their information and send randomized information to the information beneficiary.
- The receiver rebuilds the primary distribution of the data by using a

distribution reconstruction algorithm. Information got by every client is adjusted and if the quantity of clients expanded, the total data of these clients can be assessed with a great quantity of precision.

The noise parts got freely of information. The original values can't be recovered, yet the conveyance of the first record esteems can be recovered. Laterally these lines, if A is the arbitrary variable representing the information circulation for the first record, B is their regular variable representing to the commotion appropriation, and C is the arbitrary variable speaking to the last record, at that point we have: $C = A + B$, $A = C - B$.

Typically, the supposition taken that the fluctuation of the noise included is in bounteous with the goal that the first record esteems can't be effectively made a decision from the misshaped information. In Randomized reaction, the information is in such a structure, that the focal spot can't tell with probabilities superior to a pre-characterized limit whether the information from a client contains honest data or bogus data. Notwithstanding, data from every individual is disarranged, in the event that the quantity of clients is remarkably enormous, at that point the entire data of these clients can be evaluated with legitimate precision. Such property is helpful for choice tree characterization since choice tree order depends on total estimations of an informational index, instead of individual information things.

Advantages of Randomized response model are

- Randomized response based PPDM is extremely easy which does not need the information of other records in the data.
- Noise is autonomous of data
- Do not necessitate complete dataset for perturbation
- Can be applied to data compilation time
- Does not have to necessitate a reliable server with all the records to achieve anonymization procedure.
- Very basic and doesn't require information on dispersions of different records in the information.

Limitations

- Diminishing the utility of the primary data for the mining point of view.
- High altitude of individual's information loss.
- Not appropriate for various attribute database.
- Randomized response based PPDM Considers all the records equal regardless of their local density.

E. Condensation Based Model

Condensation based technique develops compelled clusters in the dataset and then produces fake data from the statistics of the constructed clusters. It generates collections of non-homogeneous size from the data, with the end goal that it is unequivocal that all record lies in a cluster whose size is at minimum identical to its anonymity intensity. Condensation based technique uses a method that summarizes the data into indefinite clusters of a predefined size, for each cluster, certain statistics are sustained. Each cluster has a size no not as much as N, which is alluded to as the degree of that privacy-preserving tactic [19] – [21]. As the level builds, the degree of privacy likewise increments. All the while, a portion of data misfortune is there as it includes the build-up of a bigger amount of records into a lone measurable gathering substance. We exploit the statistics from each gathering so as to create the relating pseudo-information.

Advantages:

- This technique utilizes pseudo-information as conflicting to changes of primary data, which assist in the enhanced conservation of privacy.
- This technique can be proficiently used for classification problems and in instance of

data streams, where data is enormously active.
- Condensation based technique is better as match up toother technologies as it utilizes pseudo-data relatively than altered data.
- There is not any compelling reason to overhaul the data mining algorithms since pseudo information is having a similar organization as of original data.

Limitations
- Data mining end results get predisposed as a raft of data misfortune is there due to the build-up of a bigger number of records into a solitary measurable gathering element.
- High level of information loss.

## V. POST PROCESSING

Privacy-preserving practices ought to fulfill a significant property known as post-processing, any post-processing of the yield of Privacy-preserving practices ought not to release any extra data information individuals. Such property has been a prerequisite that privacy definitions ought to fulfill; nonetheless, many existing privacy practices neglect to. Fig.4. provides an idea of post-processing. privacy practices only guarantee privacy of the yields of anonymization approaches and not the consequences of the postprocessing on these yields

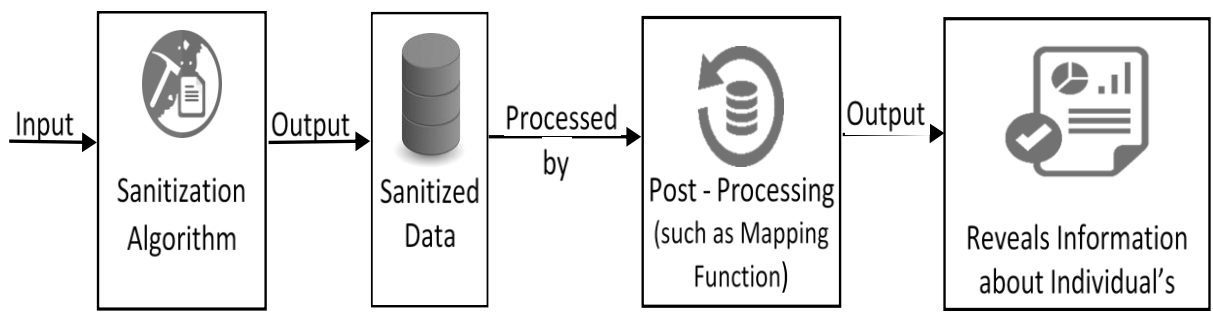

Fig. 4. Post Processing Process

Expect that the yields of anonymization approaches are prepared by the activity of post-processing, for example, a mapping capacity, to create the yields in another area. In such a circumstance, the rival is as yet incapable to separate sensitive information regardless of whether the foe knows the yields of the anonymization approaches and the yields of the post-handling.

## VI. CONCLUSION

PPDM intends to shield the secrecy of the data owner's sensitive data. while data owner's achievebeneficial information from the complete dataset. Obtaining both privacy and data utility is a great challenge. If focusing on Data Privacy then data utility or performance of mining results will be degraded. vice versa, if focusing on data utility then the privacy of the primary data will be lost. To steadinessamong the data utility and data privacy will be more challenge and it leads to many researchers to focus on this domain.

## AUTHORS PROFILE

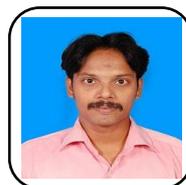

Mr. D. Dhinakaran (M'84), was born in Chennai, Tamilnadu, India, in 1984. He received the B.E degree in CSE from Anna University, Chennai, in 2006, the M.E degree in the field of CSE from Anna University, Trichy, in 2009. Presently, he is on the role of Assistant

Professor in the Department of Information Technology at Velammal Institute of Technology, Chennai. At present seeking after a Ph.D. and started exploring the field of Privacy-Preserving Data Mining, He likewise inspired by Big Data, Internet of Things, Cloud Computing, Image Processing and so forth., and he presented more than 15 papers in different National and International Conferences. He owns membership from different professional societies like IAEST, IFERP, and CSTA. He owns publications of 13 different International Journals.

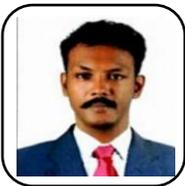

Dr. P. M. Joe Prathap (M'82). He received his B.E degree in CSE from St. Xavier's Catholic College of Engineering, Chunkankadai, in 2003 and M.E degree in the field of CSE from Karunya Institute of Technology, Coimbatore, in 2005. He received his Ph.D. degree from Anna University, Chennai, in 2011. Presently, he is on the role of Associate Professor in the Department of Information Technology at RMD Engineering College, Kavaraipettai, Tiruvallur, India. His zones of intrigue incorporate Computer Networks, Network Security, Operating Systems, Mobile Communication, and Data Mining. He has published 35 papers in various International Journals and Conferences. He has gone to numerous workshops & FDPs sponsored by AICTE; DST & IEEE related to his area of interest. He is a recognized research supervisor of Anna University, St. Peters's University, Sathyabama University &Karpagam University. He had already produced 3 Ph.D. scholars under his guidance, and as of now guiding 13 Ph.D. researchers. He is a life member of ISTE & IAENG.